\documentclass[preprint,showpacs,showkeys]{revtex4-2}
\usepackage{amsmath}
\usepackage{amssymb}

\begin{document}

\title{Gravity-induced emergence of the Fermi scale in quantum quadratic gravity}

\author{Mohammad Mehrafarin}
\email{mehrafar@aut.ac.ir}
\affiliation{Physics Department, Amirkabir University of Technology, Tehran 15916, Iran}

\begin{abstract}
In the framework of asymptotic safety, we study quantum quadratic gravity in the presence of the Higgs field considered as non-separable from the vacuum. The theory flows to a high energy fixed point where the Higgs field is strongly coupled to gravity, its potential is symmetric, and the  quadratic Weyl curvature coupling is large. The latter renders the ghost graviton an unstable high mass resonance which renders unitarity in the spirit of Lee-Week type theories. Furthermore, if the scalar graviton is tachyonic then there will be a low energy fixed point where tachyonic condensation leads to a new stable vacuum.  At this fixed point the symmetry breaks and the Fermi scale emerges, and the behavior of the Higgs field is classical (not influenced by gravitational interaction). Gravity at the UV scale is purely quadratic whereas at the Fermi scale it is linear, and  in the intermediate region both contributions are relevant. Thus, at the Fermi scale the quadratic curvature fields  disappear through the ghost instability and tachyon condensation, giving rise to Einstein gravity and the electroweak phase transition.
\end{abstract}
 

\maketitle

\section{Introduction}
A realistic quantum theory of gravity in the absence of matter and gauge fields must include the Higgs field on the grounds that it is non-separable from the vacuum. Considering quadratic gravity, our main result is that such a theory is unitary and can describe the electroweak phase transition.

Metric quantization of gravity requires higher derivatives because of renormalizability, which by producing ghost-like states conflicts with the unitarity of the theory.  For quadratic gravity, which is renormalizable \cite{Stelle1},
unitarity has been shown \cite{Donoghue} to hold  in the spirit of Lee-Wick theories \cite{Donoghue2} if the  spin-2 ghost graviton is an unstable high mass resonance in the UV scale such that the causality violation due to its backward propagation occurs on microscopic scales. Under this condition, the ghost states do not appear as asymptotic states of the scattering matrix and its unitarity is therefore preserved with no real causality concern. Resonances produce lighter modes (through decay on time scales of the order of the resonance width) which if included in the  theory assures resonance instability as required by unitarity. In vacuum, the ghost graviton of the quadratic Weyl curvature field can decay into a pair of usual massless gravitons of the linear curvature field included in the action so that unitarity then only requires a high mass for the ghost. In the context of asymptotic safety \cite{Eichhorn2, Percacci, Reichert}, this means the  theory must flow to a high energy fixed point where the  quadratic Weyl curvature coupling is large. We show that such a fixed point exists in the presence of the Higgs field  so that the theory is unitary. At this UV fixed point, which  corresponds to the strong coupling regime where the anomalous dimension of the Higgs field is large, the Higgs potential is (electroweakly) symmetric.

The coupling of the Higgs scalar with gravity shows another important feature, which is the existence of an additional (low energy) fixed point if the scalar graviton of the quadratic scalar curvature field is tachyonic. Then, the scalar graviton, which has normal forward propagation, will be inherently unstable and destabilizes the vacuum  in the high energy  phase represented by the minimum of the symmetric Higgs potential. In this interplay with the vacuum structure, tachyon condensation resolves these instabilities leading to a new stable vacuum configuration where  the Higgs field has a tachyonic mass and the symmetry spontaneously breaks. This  phase transition, which correlates with the electroweak transition, occurs  at the low energy fixed point where
the Higgs mass emerges resolving the gauge hierarchy problem \cite{Wetterich1}. Moreover, the behavior of the Higgs field is here classical and not influenced by gravitational interaction as expected. Gravity at the UV (Planck) scale is found to be purely quadratic (the linear curvature is asymptotically free) whereas at the Fermi scale it is linear (quadratic curvatures are asymptotically free), where in the intermediate region both contributions are relevant. Thus, at the Fermi scale the quadratic curvature fields completely disappear though ghost instability and tachyon condensation, giving rise to linear gravity and the massive Higgs particle. The renormalizable trajectory guides the theory from Planckian scales to the present scale via the Fermi scale, as described by the functional renormalization group (FRG) equations \cite{Reuter1, Rosten, Dupuis}.

\section{The effective action and FRG}
In the framework of asymptotic safety one works with a Euclidean action on the grounds
that metric signature is irrelevant to the fixed point behavior \cite{Manrique}. We consider the effective action involving quadratic gravity and the Higgs field $\frac{\varphi}{\surd{2}}$ (in the electroweak unitary gauge)
\begin{equation}\label{effec}
\begin{array}{l}
{\tilde\Gamma}_k =\int d^4x\, \sqrt{g}\:\bigg[\frac{1}{2} Z_k (\partial\varphi)^2+V_k(\varphi)-\frac{1}{2} M_{{\text P}k}^2(R-2\Lambda_k)
+\frac{1}{\zeta_k}\,\big(\frac{1}{2}C_{\mu\nu\rho\sigma}^2-\frac{\omega_k}{3} R^2\big) \bigg].
\end{array}
\end{equation}
Here, $k$ is the coarse-graining factor, $C_{\mu\nu\rho\sigma}$ is the Weyl tensor, and $\Lambda_k$ represents the cosmological constant. Note that a physical mass for the ghost (resp. scalar) graviton requires non-negative $\zeta_k$ (resp. $\frac{\zeta_k}{\omega_k}$), and therefore fixed points with negative $\omega_k$ pertain to tachyonic scalar graviton. The Planck mass appears in its renormalized form $M_{{\text P}k}= (8\pi G_{k})^{-1/2}$ and the  Higgs potential is $V_k(\varphi)=\frac{\mu_k^2}{2}Z_k\varphi^2+\frac{\lambda_k }{4}Z_k^2\varphi^4+\frac{\lambda_{6k}}{6}Z_k^3\varphi^6+\dots$. The couplings beyond quartic have negative mass dimension and are  irrelevant to the low energy behavior, being  relevant only in high energy scales. In practice, valuable information is obtained from suitably truncated fixed points. Such truncations are, of course, approximations because  if only one of the couplings in the potential is nonzero, then all the other ones will be generated by renormalization.

The full effective action $\Gamma_k$ in FRG  is defined by writing the average metric $g_{\mu\nu}$ in \eqref{effec} in terms of an arbitrary fixed background metric ${\hat g}_{\mu\nu}$ and the fluctuating quantum part $h_{\mu\nu}$ as $g_{\mu\nu}={\hat g}_{\mu\nu}+h_{\mu\nu}$, taking into account the classical gauge fixing action $S^{\text{gf}}[h_{\mu\nu},{\hat g}_{\mu\nu}]$ and the ensuing ghost actions  $S^{\text{gh}}_b[b_\mu, {\hat g}_{\mu\nu}],\, S^{\text{gh}}_c[c_\mu, {\bar c}_\mu, {\hat g}_{\mu\nu}]$ as in the pure gravity in the absence of the Higgs field \cite{Codello}. Thus $\Gamma_k={\tilde \Gamma}_k+S^{\text{gf}}+S^{\text{gh}}_b+S^{\text{gh}}_c$, where we refer the reader to \cite{Codello} for the explicit expressions of $S^{\text{gf}},\,S^{\text{gh}}_b$ and $S^{\text{gh}}_c$. Writing $\Gamma_k^0:=\tilde \Gamma_k+S^{\text{gf}}$, the Wetterich equation \cite{Wetterich} is
\begin{equation}\label{wett}
k\partial_k \Gamma_k=\frac{1}{2}\text {Tr} \left(\frac{k\partial_k {\cal R}_k}{\Gamma^{0(2)}_k+{\cal R}_k}\right)+T^b+T^c
\end{equation}
where $\Gamma^{0(2)}_k$ is the Hessian of $\Gamma_k^0$ in the field space of the scalar and the metric fluctuations, and the ghost traces $T^b, T^c$ are the same as in pure gravity. The Hessian and the trace are calculated in the appendices using the usual heat kernel expansion method \cite{Vassilevich, Benedetti}.

\section{The beta functions}
The LHS of the Wetterich equation can be directly calculated by differentiating \eqref{effec} (setting
 $g_{\mu\nu}={\hat g}_{\mu\nu}$) so that  the coefficients of corresponding operators on both sides can be compared.
Thus, from \eqref{final} and the results of \cite{Codello} we  obtain the gravitational beta functions 
\begin{equation}\label{beta}
\begin{array}{l}
k\partial_k {\zeta_k}=-\frac{\zeta_k^2}{60(4\pi)^2}\,(799-\eta_{\varphi k})\\
k\partial_k\omega_k=-\frac{\zeta_k}{120(4\pi)^2}\,\big[5(11-\eta_{\varphi k})+2\omega_k (1099-\eta_{\varphi k})+400\omega_k^2\big]\\
k\partial_k{\bar G}_{k}=2{\bar G}_{k}-\frac{1}{(4\pi)^2}\frac{3+26\omega_k-40\omega_k^2}{12\omega_k}\,{\bar G}_{k}\zeta_k-\frac{1}{18\pi} (\frac{3}{2}\eta_{\varphi k}
 +80+70\omega_k+8\omega_k^2)\,{\bar G}_{k}^2\\
k\partial_k{\bar \Lambda}_k=-2{\bar \Lambda}_k+
\frac{1}{256\pi(4\pi)^2}\frac{1+20\omega_k^2}{\omega_k^2}\frac{\zeta_k^2}{{\bar G}_{k}}
+\frac{1}{(4\pi)^2}\frac{1+86\omega_k+40\omega_k^2}{12\omega_k}
{\bar \Lambda}_k{\zeta_k}
-\frac{1}{(4\pi)^2}\frac{1+10\omega_k^2}{4\omega_k}\, \zeta_k\\\hspace{1.7cm}
+\frac{1}{4\pi}(8+X_k(0)) {\bar G}_{k}-\frac{1}{18\pi} (\frac{3}{2}\eta_{\varphi k} +80+70\omega_k+8\omega_k^2)\,{\bar \Lambda}_k{\bar G}_{k}
\end{array} 
\end{equation}
 (bar over parameters represents their dimensionless form) where
\begin{equation}\label{eta}
\eta_{k}=-k\partial_k \ln \surd Z_{k}= \frac{12\zeta_k(26\omega_k-1)}{\zeta_k(26\omega_k-1)-2(48\pi)^2\omega_k}
\end{equation}
is the anomalous dimension of $\varphi$ and
\begin{equation}
X_k({\varphi})=\frac{1-\frac{\eta_{k}}{3}-\frac{4}{3}\frac{\zeta_k}{\omega_k}(\omega_k+1)k^2{\bar V}_k^{\prime 2}}{1+k^2{\bar V}_k^{\prime\prime}+\frac{2}{3}\frac{\zeta_k}{\omega_k}(\omega_k+1)k^2{\bar V}_k^{\prime 2}}.
\end{equation}
(the inessential parameter $Z_k$ has been eliminated by the scalar field redefinition ${\varphi}\rightarrow \frac{\varphi}{\sqrt{Z_k}}$ so that the potential is now $\frac{\mu_k^2}{2}\varphi^2+\frac{\lambda_k }{4}\varphi^4+\frac{\lambda_{6k}}{6}\varphi^6+\dots$). Moreover, the flow of the potential is given by
\begin{equation}
\begin{array}{l}
k\partial_k {\bar V}_k=-4{\bar V}_k+\frac{1}{(4\pi)^2}\frac{\zeta_k{\bar V}_k}{1-\zeta_k{\bar V}_k}\frac{1+28\omega_k-10\zeta_k (1+\omega_k){\bar V}_k}{3\omega_k-\zeta_k(1+\omega_k){\bar V}_k}+
\frac{1}{32\pi^2}(X_k-X_k(0)).
\end{array}
\end{equation}
Differentiating both sides once (resp. twice) with respect to $\varphi^2$ and setting $\varphi=0$ yields the beta function of ${\bar \mu}_k^2$ (resp. $\lambda_k$) according to
\begin{equation}\label{beta2}
\begin{array}{l}
k\partial_k {\bar \mu}_k^2=-2{\bar \mu}_k^2+\frac{1}{(4\pi)^2}\frac{1+28\omega_k}{3\omega_k}\,\zeta_k {\bar \mu}_k^2-\frac{1}{12\pi^2}\frac{\omega_k+1}{\omega_k}\frac{\zeta_k {\bar \mu}_k^4}{1+{\bar \mu}_k^2}-\frac{1}{(4\pi)^2}\big(1-\frac{\eta_{k}}{3}\big)\frac{1}{(1+{\bar \mu}_k^2)^2}(3\lambda_k+\frac{2}{3}\frac{\omega_k+1}{\omega_k}\,\zeta_k {\bar \mu}_k^4)\\
k\partial_k {\lambda}_k=\frac{1}{(4\pi)^2}\frac{1+28\omega_k}{3\omega_k}\,\zeta_k \lambda_k+\frac{1}{(4\pi)^2}\frac{82\omega_k^2+2\omega_k+1}{9\omega_k^2}\,\zeta_k^2 {\bar \mu}_k^4+\frac{1}{6\pi^2}\frac{\omega_k+1}{\omega_k}\frac{\zeta_k{\bar \mu}_k^4}{(1+{\bar \mu}_k^2)^2}(3\lambda_k+\frac{2}{3}\frac{\omega_k+1}{\omega_k}\, \zeta_k{\bar \mu}_k^4)\\\hspace{1.6cm}-\frac{1}{3\pi^2}\frac{\omega_k+1}{\omega_k}\frac{\zeta_k \lambda_k{\bar \mu}_k^2}{1+{\bar \mu}_k^2}-
\frac{1}{8\pi^2}\big(1-\frac{\eta_{k}}{3}\big)\frac{1}{(1+{\bar \mu}_k^2)^2}(5{\bar\lambda}_{6k}+\frac{4}{3}\frac{\omega_k+1}{\omega_k}\,\zeta_k\lambda_k{\bar \mu}_k^2)\\\hspace{1.6cm}+\frac{1}{8\pi^2}\big(1-\frac{\eta_{k}}{3}\big)\frac{1}{(1+{\bar \mu}_k^2)^3}(3\lambda_k+\frac{2}{3}\frac{\omega_k+1}{\omega_k}\, \zeta_k{\bar \mu}_k^4)^2.
\end{array}
\end{equation}
Similarly, the beta function of ${\bar \lambda}_{6k}$ involves $({\bar\lambda}_{6k},{\bar \lambda}_{8k})$, and so on, such that these equations form an infinite  hierarchy. Considering sixth order truncation of the fixed point, nontrivial fixed points  will have nonzero ${\bar\lambda}_{6}$, which  yields the remaining  potential parameters through \eqref{beta2}. This  nonzero value is constrained to be positive so that the truncated potential at the high energy fixed point is bounded from below. Moreover, at the low energy fixed point where $\lambda$ is positive too, $ \lambda_{6}$ becomes irrelevant resulting in the usual lower-bounded quartic potential. 

\section{The fixed points and implications}
There exist two interacting fixed points. For the `UV'  fixed point, the first two beta functions in \eqref{beta} yield $\eta^\star=799$ and $\omega^\star\approx -3.977$   (the positive solution $\omega^\star \approx 2.477$ corresponds to a physical scalar graviton) so that \eqref{eta} then determines $\zeta^\star$. The Higgs field is hence a UV-relevant operator and the large anomalous dimension implies that its high energy behavior is drastically influenced by quantum gravity effects. At this  fixed point the gravitational parameters are
\begin{equation}
\zeta^\star\approx 1759,\quad\omega^\star\approx -3.977,\quad{\bar G}^\star=0,\quad{\bar \Lambda}^\star\approx1.591
\end {equation}
where all the parameters are UV-attractive except for $\omega_k$ which is UV-repulsive. The large value  of the coupling $\zeta$ indicates unitarity with acausality on microscopic scales. Therefore, at Planckian scales gravity is purely quadratic as the linear curvature field is asymptotically free. Then, for the electroweak parameters, based the numerical solution of \eqref{beta2} with a wide range of positive values of ${\bar\lambda}_{6}^\star$ one concludes that ${\bar \mu}^{\star 2}>0,\,\lambda^\star<0$. (Negative $\lambda$ conforms with the relevance of $\lambda_{6}$ in the potential at high energy.) The potential is therefore symmetric with zero Higgs v.e.v, as anticipated in the UV region, and we thus
 have the ghost and the tachyonic graviton fields in the stable vacuum state.

In the IR flow beyond the Planck scale, the vacuum sate becomes progressively more unstable because of the decay of the tachyonic graviton field, while the decay of the ghost graviton produces the usual massless gravitons. The flow takes the theory on the critical surface from the high energy to the low energy fixed point where
\begin{equation}
\zeta^\star=0,\quad\omega^\star\approx -0.0251,\quad{\bar G}^\star\approx 1.445,\quad{\bar \Lambda}^\star\approx 0.2636 
\end {equation}
(there is also the physically inadmissible solution $\omega^\star\approx -5.470$ which yields negative gravitational constant). Note $\eta^*=0$, so that the low energy behavior of the Higgs field is classical and not influenced by gravitational interaction. Again from \eqref{beta2}, for the electroweak parameters we have
\begin{equation}
{\bar \mu}^{\star 2}(1+{\bar \mu}^{\star 2})^2=-\frac{3}{32\pi^2}\lambda^\star,\quad\lambda^{\star 2}=\frac{5}{9}(1+{\bar \mu}^{\star 2})\,{\bar \lambda}_{6}^\star
\end {equation}
and so $-1<{\bar \mu}^{\star 2}<0$ ($\lambda^\star>0$).  Hence, tachyonic graviton condensation results in a tachyonic mass for the Higgs field and a new stable vacuum with nonzero v.e.v $\sqrt{-\mu^{ 2}/\lambda}$ corresponding to spontaneous symmetry breaking. Here, all the parameters (the irrelevant ${\lambda}_{6}$ not included) are UV-attractive except for $\zeta_k, \omega_k$ which are marginal. At the  fixed point, which correlates with the electroweak transition,  the quadratic curvature fields are asymptotically free, having given rise through the ghost instability and tachyon condensation to linear gravity and to mass for the Higgs particle, respectively.

We can obtain the flow of ${\bar G}_k, {\bar \Lambda}_k$ beyond the Fermi scale by setting the marginal parameters $\zeta_k, \omega_k$ to their fixed point values in the flow equations \eqref{beta}. Then, analytical solution yields
\begin{equation}
G_k=\frac{G_0}{1+ G_0 k^2/{\bar G}^\star},\quad \Lambda_k=\frac{\Lambda_0+G_0 k^4/2\pi}{1+ G_0 k^2/{\bar G}^\star}
\end{equation}
which hold up to the Fermi scale and the first relation can also be written as $M_{{\text P}k}^2=M_{{\text P}0}^2+\frac{k^2}{8\pi{\bar G}^\star}\approx M_{{\text P}0}^2+0.0275\, k^2$.\\

\noindent {\bf Data Availability Statement}: No data are available because of the nature of the research.

\appendix\section{Calculation of the Hessian}
Writing ${\tilde\Gamma}_k={\tilde\Gamma}^{(g)}_k+{\tilde \Gamma}^{(\varphi)}_k$, where ${\tilde\Gamma}^{(g)}_k$ is the pure gravity part and ${\tilde \Gamma}^{(\varphi)}_k$ the piece involving $\varphi$, we proceed to calculate the Hessian $\Gamma^{0(2)}_k$ of $\Gamma_k^0={\tilde\Gamma}^{(g)}_k+S^{\text{gf}}+{\tilde \Gamma}^{(\varphi)}_k$. 

For the pure gravity part $({\tilde\Gamma}^{(g)}_k+S^{\text{gf}})$ the gravitational Hessian is \cite{Shapiro} 
\begin{equation}\label{hes} 
K\, (I\,\Delta^2+ V^{\eta\lambda}{\hat \nabla_\eta}{\hat \nabla_\lambda}+U)
\end{equation}
where $\Delta=-\hat \nabla^2$, and covariant derivatives are with respect to the background metric, with which indices are also raised and lowered. Here $I$ is the unit matrix with elements $\delta^{(\rho}_{\ \ \alpha}\,\delta^{\sigma)}_{\ \beta}$, and
\begin{equation}
K^{\mu\nu}_{\ \ \rho\sigma}=\frac{1}{2\zeta_k}\,\left(\delta^{(\mu}_{\ \ \rho} \, \delta^{\nu)}_{\ \sigma} -\frac{1}{4}\,\frac{1+4\omega_k}{1+\omega_k}{\hat g}^{\mu\nu}{\hat g}_{\rho\sigma}\right).
\end{equation}
For the rather lengthy expressions of $V^{\eta\lambda}$ and $U$ in  terms of background quantities the reader is referred to \cite{Shapiro}. To calculate the gravitational Hessian of ${\tilde\Gamma}^{(\varphi)}_{k}$ we introduce $\psi_{\mu\nu}:= \frac{1}{2} Z_{ k} \partial_\mu \varphi \partial_\nu \varphi$, so that 
\begin{equation}\label{var}
\begin{array}{l}
{\tilde\Gamma}^{(\varphi)}_k[{\hat g}_{\mu\nu}+{h}_{\mu\nu}, \varphi]=\\\int d^4x \sqrt{{\hat g}}\,\big(1+\frac{1}{2} h-\frac{1}{4}{ h}_{\mu\nu}J^{\mu\nu}_{\ \ \rho\sigma}{h}^{\rho \sigma}\big)\left[({\hat g}^{\alpha\beta}-{ h}^{\alpha\beta}+{ h}^\alpha_{\ \gamma}{ h}^{\gamma\beta}) \psi_{\alpha\beta}+V_k(\varphi)\right]
\end{array}
\end{equation}
having defined
$$
J^{\mu\nu}_{\ \ \rho\sigma}:=\delta^{(\mu}_{\ \ \rho} \, \delta^{\nu)}_{\ \sigma} -\frac{1}{2}\,{\hat g}^{\mu\nu}{\hat g}_{\rho\sigma}.
$$
The term in \eqref{var} quadratic in ${ h}_{\mu\nu}$ reads as
\begin{equation}
{\tilde\Gamma}^{(\varphi)\text{quad}}_k=-\frac{1}{4}\int d^4x \sqrt{{\hat g}}\,{ h}_{\mu\nu}\left[J^{\mu\nu}_{\ \ \alpha\beta}(\psi^\sigma_{\ \sigma}+V_k)+{\hat g}^{\mu\nu}\psi_{\alpha\beta}+{\hat g}_{\alpha\beta}\psi^{\mu\nu} -4\,\delta^{(\mu}_{\ (\alpha} \, \psi^{\nu)}_{\ \beta)}\right]\,{ h}^{\alpha\beta}
\end{equation}
so that the Hessian is
\begin{equation}
-\frac{1}{2}\,\left[J^{\mu\nu}_{\ \ \alpha\beta}(\psi^\sigma_{\ \sigma}+V_k)+{\hat g}^{\mu\nu}\psi_{\alpha\beta}+{\hat g}_{\alpha\beta}\psi^{\mu\nu} -4\,\delta^{(\mu}_{\ (\alpha} \, \psi^{\nu)}_{\ \beta)}\right].
\end{equation}
We write this in obvious notation as ${\tilde\Gamma}^{(\varphi)(2)}_{k,\text {grav}}=-\frac{1}{2}\,[J(\psi+V_k)+T]$, which also defines $T$. A quick calculation shows that $J=KL$, where 
\begin{equation}
L^{\rho\sigma}_{\ \ \alpha\beta}=2\zeta_k\,\left(\delta^{(\rho}_{\ \ \alpha} \, \delta^{\sigma)}_{\ \beta} +\frac{1-2\omega_k}{12\omega_k}{\hat g}^{\rho\sigma}{\hat g}_{\alpha\beta}\right).
\end{equation}
Hence, we can factorize $K$ and wrie the Hessian as $KW$, where $W=-\frac{1}{2}\,[L(\psi+V_k)+K^{-1}T]$, which must be added to \eqref{hes} to get the gravitational Hessian of $\Gamma_k^0$. $K^{-1}$ is the inverse of $K$ ($KK^{-1}=I$) which is given by
\begin{equation}
\left(K^{-1}\right)^{\rho\sigma}_{\ \ \mu\nu}=2\zeta_k\,\left(\delta^{(\rho}_{\ \ \mu} \, \delta^{\sigma)}_{\ \nu} -\frac{1+4\omega_k}{12\omega_k}{\hat g}^{\rho\sigma}{\hat g}_{\mu\nu}\right).
\end{equation}
Evaluating $K^{-1}T$ thus yields
\begin{equation}
\begin{array}{l}
W^{\rho\sigma}_{\ \ \, \alpha\beta}=-\zeta_k\,\bigg[J^{\rho\sigma}_{\ \ \alpha\beta}\psi^\mu_{\ \mu}+{\hat g}^{\rho\sigma}\psi_{\alpha\beta}+{\hat g}_{\alpha\beta} \psi^{\rho\sigma}-4\delta^{(\rho}_{\ ( \alpha} \, \psi^{\sigma)}_{\ \beta)} 
+\left(\delta^{(\rho}_{\ \ \alpha} \, \delta^{\sigma)}_{\ \beta}+\frac{1-2\omega_k}{12\omega_k}{\hat g}^{\rho\sigma}{\hat g}_{\alpha\beta}\right) V_k\bigg].
\end{array}
\end{equation}
Therefore,  the gravitational Hessian is
\begin{equation}\label{gravh}
\Gamma^{0(2)}_{k,\text {grav}}=K\, (I\,\Delta^2+ V^{\eta\lambda}{\hat \nabla_\eta}{\hat \nabla_\lambda}+U+W).
\end{equation}

The complete Hessian is obtained  in the field space of the scalar and the metric fluctuations by letting $\varphi \rightarrow \varphi+\phi$ in \eqref{var}, as the block matrix
\begin{equation}
\Gamma^{0(2)}_k=
\left(
\begin{array}{cc}
\Gamma^{0(2)}_{k,\varphi}&N_j\\ N^{i\dagger }&(\Gamma^{0(2)}_{k,\text {grav}})^i_{\ j}
\end{array}
\right)
\end{equation}
where we have abbreviated pairs of external symmetric indices with a single index (e.g. $N_j\equiv N_{\mu\nu}, \, K^i_{\ j}\equiv K^{\rho\sigma}_{\ \ \mu\nu}$). Here
\begin{equation}\label{14}
\Gamma^{0(2)}_{k,\varphi}=Z_{ k}\Delta+V_k^{\prime\prime}(\varphi)
\end{equation}
 is the Hessian with respect to $\varphi$ and $N_j$ is the mixed Hessian given by (${\hat\nabla}_\beta^\dagger=-{\hat\nabla}_\beta$)
\begin{equation}\label{15}
N_{\mu\nu} = Z_{k}  J^{\alpha\beta}_{\ \ \mu\nu}\,(\partial_\alpha\varphi{\hat\nabla}_\beta+\frac{1}{2}{\hat\nabla}_\alpha{\hat\nabla}_\beta\varphi)+\frac{1}{2} {\hat g}_{\mu\nu} V_k^\prime.
\end{equation}

\section{Calculation of the trace}
The regulator ${\cal R}_k$ is diagonal in the field space so that 
\begin{equation}
\Gamma^{0(2)}_k+{\cal R}_k=
\left(
\begin{array}{cc}
\Gamma^{0(2)}_{k,\varphi}+{\cal R}_k^\varphi&N_j\\ N^{i\dagger}&(\Gamma^{0(2)}_{k,\text {grav}}+{\cal R}_{k}^{\text{grav}})^i_{\ j}
\end{array}
\right).
\end{equation}
We choose 
${\cal R}_k^{\text{grav}}(\Delta)=k^4P(\Delta^2/k^4)K$ as in \cite{Codello} (for the use of another type of cut-off in the gravitational sector see \cite{Groh}), and ${\cal R}_k^\varphi (\Delta)=Z_{k}\,k^2P(\Delta/k^2)$, where  $P(y)=(1-y) \Theta(1-y)$ is the optimal cut-off function ($\Theta$ is the unit-step function). Since $k\partial_k {\cal R}_k$ is also diagonal, for the trace we need only the diagonal elements of $(\Gamma^{0(2)}_k+{\cal R}_k)^{-1}$, namely
\begin{eqnarray}\label{diagtr}
(\Gamma^{0(2)}_k+{\cal R}_k)^{-1}_{\varphi\varphi}=[(\Gamma^{0(2)}_{k,\varphi}+{\cal R}_k^\varphi)-N_i\, (\Gamma^{0(2)}_{k,\text {grav}}+{\cal R}_{k}^{\text{grav}})^{-1i}_{\ \ \ \,j}N^{\dagger j}]^{-1}  \nonumber \\
(\Gamma^{0(2)}_k+{\cal R}_k)^{-1i}_{hh\ j}=[(\Gamma^{0(2)}_{k,\text {grav}}+{\cal R}_k^{\text{grav}})^i_{\ j}-N^{\dagger\,i}\,(\Gamma^{0(2)}_{k,\varphi}+{\cal R}_k^\varphi)^{-1}N_j]^{-1}.
\end{eqnarray}
We therefore have
\begin{equation}\label{trr}
\frac{1}{2}\text{Tr}\left(\frac{k\partial_k{\cal R}_k}{\Gamma^{0(2)}_k+{\cal R}_k}\right)=\frac{1}{2}\text{Tr}\left(\frac{k\partial_k{\cal R}_k}{\Gamma^{0(2)}_k+{\cal R}_k}\right)_{\varphi\varphi}+\frac{1}{2}\text{Tr}\left(\frac{k\partial_k{\cal R}_k}{\Gamma^{0(2)}_k+{\cal R}_k}\right)_{hh}.
\end{equation}

 Now, using \eqref{gravh}, \eqref{14}, \eqref{15} and ignoring all operators that do not contribute to the flow according to our effective action, 
\begin{equation}
\begin{array}{l}
T^\varphi=\frac{1}{2}\text{Tr}\left(\frac{k\partial_k{\cal R}_k}{\Gamma^{0(2)}_k+{\cal R}_k}\right)_{\varphi\varphi}\\ =\text{Tr}\,\bigg([k^2(1-\eta_{k})P_1-\Delta P_1^\prime]\left[\Delta+k^2 P_1+\frac{V_k^{\prime\prime}}{Z_{ k}}+\frac{2}{3}\frac{\zeta_k}{\omega_k}(1+\omega_k)\frac{V_k^{\prime 2}}{Z_{ k}(\Delta^2+k^4P_2)}
-\frac{O^{\eta\lambda}{\hat \nabla_\eta}{\hat \nabla_\lambda}}{\Delta^2+k^4P_2}\right]^{-1}\bigg)\\  
=\text{Tr}\,\bigg(\frac{k^2(1-\eta_{k})P_1-\Delta P_1^\prime}{\Delta+k^2 P_1}\\ \hspace{2cm} \big[1+\Sigma_{n=1}^\infty\big(\frac{-1}{Z_{k}(\Delta+k^2 P_1)}\big)^n\big(V_k^{\prime\prime}+\frac{2}{3}\frac{\zeta_k}{\omega_k}(1+\omega_k)\frac{V_k^{\prime 2}}{\Delta^2+k^4P_2}\big)^n +\frac{O^{\eta\lambda}{\hat \nabla_\eta}{\hat \nabla_\lambda}}{(\Delta+k^2 P_1)(\Delta^2+k^4P_2)}\big]\bigg)
\end{array}
\end{equation}
where $P_1\equiv P(\Delta/k^2),\,P_2\equiv P(\Delta^2/k^4)$, prime denotes derivative with respect to argument, $\eta_{ k}=-k\partial_k \ln \surd{Z_{k}}$ is the anomalous dimension of $\varphi$, and 
\begin{equation}
O^{\eta\lambda}=-Z_{k} L^{\alpha\eta}_{\ \ \,\mu\nu}J^{\beta\lambda\mu\nu}\, \partial_\alpha \varphi \, \partial_\beta \varphi=- Z_{ k}(K^{-1})^{\alpha\eta\beta\lambda}\,\partial_\alpha \varphi\, \partial_\beta\varphi.
\end{equation}
Thus
\begin{equation}\label{sctr}
\begin{array}{l}
T^\varphi=\text{Tr}(F_{0,0}(\Delta))+ \Sigma_{n=1}^\infty(-1)^n \big\{a^n\text{Tr}(F_{n,0}(\Delta))+na^{n-1}b\,\text{Tr}(F_{n,1}(\Delta))\\ \hspace{1cm}+\frac{n(n-1)}{2}a^{n-2}b^2\,\text{Tr}(F_{n,2}(\Delta))+\dots +b^n\, \text{Tr}(F_{n,n}(\Delta))\big\}
+\text{Tr}\big(F_{1,1}(\Delta) O^{\eta\lambda}{\hat \nabla_\eta}{\hat \nabla_\lambda}\big)
\end{array}
\end{equation}
where $a\equiv \frac{V_k^{\prime\prime}}{Z_{ k}},\, b\equiv \frac{2}{3}\frac{\zeta_k}{\omega_k}(1+\omega_k)\frac{V_k^{\prime 2}}{Z_{k}}$, and 
\begin{equation}
F_{n,m}(\Delta)=\frac{k^2(1-\eta_{k})P_1-\Delta P_1^\prime}{(\Delta+k^2 P_1)^{n+1}(\Delta^2+k^4P_2)^m}.
\end{equation}
The trace  \eqref{sctr} is evaluated to be
\begin{equation}
\begin{array}{l}
T^\varphi=\frac{1}{(4\pi)^2} \int d^4x \sqrt{{\hat g}}\\ \hspace{1cm} \bigg \{Q_2[F_{0,0}]+\Sigma_{n=1}^\infty (-1)^n \big(a^n Q_2[F_{n,0}]+na^{n-1}b\,Q_2[F_{n,1}]+\dots+b^n\,Q_2[F_{n,n}]\big)\\ \hspace{1.5cm}-\frac{1}{2}Q_3[F_{1,1}]  O^\eta_{\ \eta}
+\frac{1}{6}{\hat R}\,Q_1[F_{0,0}]+\frac{1}{360}(3{\hat C}_{\mu\nu\rho\sigma}^2+5{\hat R}^2)F_{0,0}(0)\bigg \} 
\end{array}
\end{equation}
 where 
\begin{equation}\label{Q}
Q_m[f(y)]=\frac{1}{\Gamma(m)}\int_0^\infty dy\, y^{m-1} f(y). 
\end{equation}
Hence  
\begin{equation}
Q_1[F_{0,0}]=k^2 \big(1-\frac{\eta_{ k}}{2}\big),\
Q_2[F_{n,m}]=\frac{1}{2} k^{4-2n-4m}\big(1-\frac{\eta_{ k}}{3}\big),\
 Q_3[F_{1,1}]=\frac{1}{6}\big(1-\frac{\eta_{ k}}{4}\big)
\end{equation}
which yields  (bar over parameters represents their dimensionless form)
\begin{equation} 
\begin{array}{l}
T^\varphi=\frac{1}{(4\pi)^2} \int d^4x \sqrt{{\hat g}}\, \bigg [\frac{1}{72}\frac{\zeta_k}{\omega_k}(26\omega_k-1)\big(1-\frac{\eta_{ k}}{4}\big)Z_{k}(\partial\varphi)^2 \\\hspace{1cm}
 +\frac{k^4}{2}\frac{1-\frac{\eta_{ k}}{3}}{1+k^2\frac{{\bar V}_k^{\prime\prime}}{Z_{ k}}+\frac{2}{3}\frac{\zeta_k}{\omega_k}(\omega_k+1)k^2\frac{{\bar V}_k^{\prime 2}}{Z_{k}}} 
+\frac{k^2}{6}\big(1-\frac{\eta_{ k}}{2}\big)\,{\hat R}+\frac{1-\eta_{ k}}{360}(3{\hat C}_{\mu\nu\rho\sigma}^2+5{\hat R}^2)\bigg].
\end{array}
\end{equation}
Similarly
\begin{equation}\label{grav}
\begin{array}{l}
\frac{1}{2}\text{Tr}\left(\frac{k\partial_k{\cal R}_k}{\Gamma^{0(2)}_k+{\cal R}_k}\right)_{hh}=2\,\text{Tr}\,\bigg((k^4P_2-\Delta^2P_2^\prime)\\\left[I(\Delta^2+k^4P_2)+V^{\eta\lambda}{\hat \nabla_\eta}{\tilde \nabla_\lambda}+U+W+\frac{2}{3}\frac{\zeta_k}{\omega_k}(1+\omega_k)\frac{V_k^{\prime 2}S}{Z_{ k}(\Delta+k^2P_1)+V_k^{\prime\prime}}-\frac{S^{\eta\lambda}{\hat \nabla_\eta}{\hat \nabla_\lambda}}{\Delta+k^2 P_1}\right]^{-1}\bigg)\\
=2\,\text{Tr}\,\bigg(\frac{k^4P_2-\Delta^2P_2^\prime}{\Delta^2+k^4P_2}\bigg[I-\frac{V^{\eta\lambda}{\hat \nabla_\eta}{\hat \nabla_\lambda}+U}{\Delta^2+k^4P_2}+\left(\frac{V^{\eta\lambda}{\hat \nabla_\eta}{\hat \nabla_\lambda}}{\Delta^2+k^4P_2}\right)^2+\Sigma_{n=1}^\infty \left(\frac{\zeta_k V_k}{\Delta^2+k^4P_2}\right)^n H_n\\\hspace{3.5cm}
+\Sigma_{n=1}^\infty \frac{(-b)^n S}{(\Delta^2+k^4P_2)^n(\Delta+k^2P_1+a)^n}+\frac{S^{\eta\lambda}{\hat \nabla_\eta}{\hat \nabla_\lambda}}{(\Delta^2+k^4P_2)(\Delta+k^2 P_1)}\bigg]\bigg)
\end{array}
\end{equation}
(the $k$-dependence of all couplings in the regulator are neglected which corresponds to the one-loop approximation of the RG equations) where   
\begin{equation}
H_n\equiv \big(I+\frac{1-2\omega_k}{3\omega_k}\, S\big)^n=I+\big[\big(\frac{1+\omega_k}{3\omega_k}\big)^n-1\big] S, \quad S^{\mu\nu}_{\ \ \,  \rho\sigma}=\frac{1}{4} {\hat g}^{\mu\nu} {\hat g}_{\rho\sigma}
\end{equation}
comes from $(-W)^n$, and
\begin{equation}
 (S^{\eta\lambda})_{\mu\nu \rho\sigma}=-Z_{ k} L^{\alpha\eta}_{\ \ \,\mu\nu}J^{\beta\lambda}_{\ \ \, \rho\sigma}\, \partial_\alpha \varphi \, \partial_\beta \varphi
\end{equation}
(note that $S=S^2=S^3=\dots$ and $\text{Tr} (-W)=\text{Tr}(\zeta_k V_k H_1)$). The first three terms in \eqref{grav} give the trace in pure gravity in the absence of the Higgs field, which, when added to the two ghost traces $T^b, T^c$ yields the result obtained in \cite{Codello}. Thence, the contribution of the Higgs field to the above trace is given by
\begin{equation}\label{trphi}
T^{(\varphi)}_\text{grav}=2\, \Sigma_{n=1}^\infty (\zeta_k V_k)^n\text{Tr}\left(g_{n,0}(\Delta) H_n\right)+2\, \Sigma_{n=1}^\infty (-b)^n\text{Tr}\left(f_{n,n}(\Delta,a) S\right)+2\, \text{Tr}\big(g_{1,1}(\Delta) S^{\eta\lambda}{\hat \nabla_\eta}{\hat \nabla_\lambda}\big)
\end{equation}
where 
\begin{equation}
f_{n,m}(\Delta,a)=
\frac{k^4P_2-\Delta^2P_2^\prime}{(\Delta^2+k^4P_2)^{n+1}(\Delta+k^2 P_1+a)^m},\ g_{n,m}(\Delta)=f_{n,m}(\Delta,0).
\end{equation}
Thus, from \eqref{wett} and \eqref{trr}, adding $T^{(\varphi)}_\text{grav}+T^\varphi$ to  the result of \cite{Codello} yields $k\partial_k \Gamma_k$.
Calculating  \eqref{trphi}, we have
\begin{equation}
\begin{array}{l}
T^{(\varphi)}_\text{grav}=\frac{2}{(4\pi)^2} \int d^4x \sqrt{{\hat g}}\\ \{\Sigma_{n=1}^\infty(\zeta_k V_k)^n Q_2[g_{n,0}]\, (H_n)^{\mu\nu}_{\ \ \, \mu\nu}+ \Sigma_{n=1}^\infty (-b)^n Q_2[f_{n,n}]\,S^{\mu\nu}_{\ \ \, \mu\nu}
-\frac{1}{2}Q_3[g_{1,1}]\, (S^\eta_{\ \eta})^{\mu\nu}_{\ \ \mu\nu}\}
\end{array}
\end{equation}
where $Q_2[g_{n,0}]=\frac{1}{2}k^{4-4n},\,Q_2[f_{n,n}]=\frac{1}{2}\frac{1}{(k^2+a)^n}k^{4-4n},\, Q_3[g_{1,1}]=\frac{1}{6}$ so that
\begin{equation}\label{trace phi grav}
\begin{array}{l}
T^{(\varphi)}_\text{grav}=\frac{1}{(4\pi)^2} \int d^4x \sqrt{{\hat g}}\big[\frac{1}{36}\frac{\zeta_k}{\omega_k}(26\omega_k-1)\,Z_{ k}(\partial\varphi)^2+\frac{\zeta_k V_k}{1-\zeta_k{\bar V}_k}\frac{1+28\omega_k-10\zeta_k (1+\omega_k){\bar V}_k}{3\omega_k-\zeta_k(1+\omega_k){\bar V}_k}\\\hspace{4cm}-k^4\frac{\frac{2}{3}\frac{\zeta_k}{\omega_k}(1+\omega_k)k^2\frac{{\bar V}_k^{\prime 2}}{Z_{k}}}{1+k^2\frac{{\bar V}_k^{\prime\prime}}{Z_{ k}}+\frac{2}{3}\frac{\zeta_k}{\omega_k}(1+\omega_k)k^2\frac{{\bar V}_k^{\prime 2}}{Z_{ k}}}\big].
\end{array}
\end{equation}
Therefore, collecting results for $T^\varphi$ and $T^{(\varphi)}_\text{grav}$, we have
\begin{equation}\label{final}
\begin{array}{l}
T^{(\varphi)}_\text{grav}+T^\varphi=\frac{1}{(4\pi)^2} \int d^4x \sqrt{{\hat g}}\,\big[\frac{1}{24}\frac{\zeta_k}{\omega_k}(26\omega_k-1) \big(1-\frac{\eta_{k}}{12}\big)\,Z_{ k}(\partial\varphi)^2 +\frac{k^4}{2} X_k(0)\\  \hspace{4cm}+\frac{\zeta_k V_k}{1-\zeta_k{\bar V}_k}\frac{1+28\omega_k-10\zeta_k (1+\omega_k){\bar V}_k}{3\omega_k-\zeta_k(1+\omega_k){\bar V}_k} 
+\frac{k^4}{2}(X_k-X_k(0))
\\\hspace{4cm}
+\frac{k^2}{6} \big(1-\frac{\eta_{ k}}{2}\big)\,{\hat R}+\frac{1}{360}(1-\eta_{ k}) (3{\hat C}_{\mu\nu\rho\sigma}^2+5{\hat R}^2)\big]
\end{array}
\end{equation}
where
\begin{equation}
X_k(\varphi)=\frac{1-\frac{\eta_{ k}}{3}-\frac{4}{3}\frac{\zeta_k}{\omega_k}(\omega_k+1)k^2\frac{
{\bar V}_k^{\prime 2}}{Z_{k}}}
{1+k^2\frac{{\bar V}_k^{\prime\prime}}{Z_{ k}}+\frac{2}{3}\frac{\zeta_k}{\omega_k}(\omega_k+1)k^2\frac{{\bar V}_k^{\prime 2}}{Z_{ k}}}.
\end{equation}
 In \eqref{final}, $X_k(0)$ contributes to the flow of the cosmological constant and the second line (which vanishes in the absence of potential) gives the flow of the potential.


\begin{thebibliography}{widest-label}
\bibitem{Stelle1} K. S. Stelle, Phys. Rev. D {\bf 16}, 953 (1977).
\bibitem{Donoghue} J. F. Donoghue and G. Menezes, Phys. Rev. D {\bf 100}, 105006 (2019).
\bibitem{Donoghue2} J. F. Donoghue and G. Menezes, Phys. Rev. D {\bf 99}, 065017 (2019).
\bibitem{Eichhorn2} A. Eichhorn, Front. Astron. Space Sci. {\bf 5}, 47 (2019).
\bibitem{Percacci} R. Percacci, {\it An introduction to Covariant Quantum Gravity and Asymptotic Safety} (World Scientific, 2017).
\bibitem{Reichert} M. Reichert, Proc. Sci. {\bf 384}, 005 (2020).
\bibitem{Wetterich1} C. Wetterich and M. Yamada, Phys. Lett. B {\bf 770}, 268 (2017).
\bibitem{Reuter1} M. Reuter and F. Saueressig, {\it Quantum Gravity and Functional Renormalization Group: The road towards Asymptotc Safety} (Cambridge University Press, 2019).
\bibitem{Rosten} O. J. Rosten, Phys. Rep. {\bf 511}, 177 (2012).
\bibitem{Dupuis} N. Dupuis, L. Canet, A. Eichhorn, W. Metzner, J. M. Pawlowski, M. Tissier, and N. Wschebor, Phys. Rep. {\bf 910}, 1 (2021).
\bibitem{Manrique} E. Manrique, S. Rechenberger, and F. Saueressig, Phys. Rev. Lett. {\bf 106}, 251302 (2011).
\bibitem{Codello} A. Codello and R. Percacci, Phys. Rev. Lett. {\bf 97}, 221301 (2006).
\bibitem{Wetterich} C. Wetterich, Phys. Lett. B {\bf 301}, 90 (1993).
\bibitem{Vassilevich} D. V. Vassilevich, Phys. Rep. {\bf 388}, 279 (2003).
\bibitem{Benedetti} D. Benedetti, K. Groh, P. F. Machado, and F. Sauerissig, JHEP {\bf 2011}, 79 (2011).
\bibitem{Shapiro} G. de Berredo-Peixoto and I. L. Shapiro, Phys. Rev. D {\bf 71}, 064005 (2005).
\bibitem{Groh} K. Groh, S. Rechenberger, F. Saueressig, and O. Zanusso, Proc. Sci. {\bf 134}, 124 (2012).
\end{thebibliography}
\end{document}